\documentclass[trackchanges]{aastex7}
\usepackage{amsmath}

\usepackage{verbatim}
\usepackage{hyperref}
\usepackage{kotex}
\usepackage{natbib}

\DeclareUnicodeCharacter{039B}{\ensuremath{\Lambda}}

\newcommand{\z}[1]{{\textcolor{red}{#1}}}

\begin{document}

\title{Living with Neighbors. VI. Unraveling the Dual Impact of Bars on Star Formation in Paired Galaxies Using DESI} 

\author[orcid=0000-0003-0960-687X, sname='Zee']{Woong-Bae G. Zee}
\affiliation{School of Liberal Studies, Sejong University, Seoul, 05006, Republic of Korea}
\email[show]{galaxy.wb.zi@gmail.com}  

\author[orcid=0000-0002-1842-4325, sname='Yoon']{Suk-Jin Yoon}
\affiliation{Department of Astronomy, Yonsei University, Seoul, 03722, Republic of Korea}
\affiliation{Center for Galaxy Evolution Research, Yonsei University, Seoul, 03722, Republic of Korea}
\email[show]{sjyoon0691@yonsei.ac.kr}

\begin{abstract}

We present a comprehensive investigation into the influence of stellar bars on star formation (SF) in galaxy pairs, using a large sample of low-redshift galaxies ($0.02$\,$<$\,z\,$<$\,$0.08$) from the DESI Legacy Imaging Surveys DR8. 
Our analysis examines whether bars enhance or suppress SF during pair interactions, and how these outcomes depend on the star-forming properties of companion galaxies. 
We find that bars either catalyze or inhibit SF in their host galaxies, depending on the companion's SF activity. 
In particular, barred galaxies paired with actively star-forming companions experience more pronounced central starbursts (with sSFR up to $\sim$\,2.5 dex higher) than unbarred counterparts, whereas those with passive companions often have suppressed SF (sometimes below isolated galaxy levels).
The notion of the dual role of bars can reconcile conventional conflicting reports of bar-driven enhancement versus quenching of SF activity.
Bars, well known to regulate kpc-scale dynamics, may also link to the impact of external environments: when a star-forming companion provides sufficient gas, bars drive central starbursts, whereas in gas-poor interactions, bars hasten gas depletion and contribute to SF suppression. 
This work highlights the necessity of accounting for both internal structure and companion properties to fully understand SF regulation in interacting galaxies. 

\end{abstract}


\section{Introduction} \label{sec:intro}

Galaxy interactions are a well-established mechanism for triggering enhanced star formation (SF). 
Early simulations by \citet{1972ApJ...178..623T} demonstrated that tidal forces during encounters can funnel gas toward galactic centers, setting the stage for later models where such inflows ignite central starbursts (e.g., \citealt{1991ApJ...370L..65B}; \citealt{1992ApJ...387..152J}; \citealt{1996ApJ...464..641M}). 
Observationally, interacting and morphologically disturbed galaxies exhibit elevated star formation rates (SFRs) and efficiencies (SFEs), often enhanced by factors of 2$-$3 compared to isolated systems (e.g., \citealt{1978ApJ...219...46L}; \citealt{1985AJ.....90..708K}; \citealt{1987AJ.....93.1011K}; \citealt{1988ApJ...325...74S}; \citealt{1995A&A...304..325G}; \citealt{2004AJ....128.2170H}).
Hydrodynamical simulations confirmed that major mergers drive gas inflows and nuclear starbursts (e.g., \citealt{2008MNRAS.391.1137L}; \citealt{2011MNRAS.415.3750M}; \citealt{2012ApJ...746..108T}; \citealt{2020MNRAS.494.4969P}; \citealt{2023MNRAS.518.3261P}). 
Parallel observational studies, particularly using the Sloan Digital Sky Survey (SDSS), showed that galaxies in close pairs exhibit higher central SFRs and gas concentrations than isolated counterparts (\citealt{2002ApJS..143...47D}; \citealt{2008MNRAS.385.1915L}; \citealt{2015MNRAS.454.1742K}; \citealt{2025ApJ...982..130P}).

Large statistical surveys further demonstrated that SF is systematically enhanced in close galaxy pairs (e.g., \citealt{2008AJ....135.1877E}; \citealt{2011MNRAS.412..591P}; \citealt{2013MNRAS.433L..59P}; \citealt{2021ApJ...909..120S}). 
For instance, \citet{2004MNRAS.355..874N} showed that specific SFRs (sSFRs) increase significantly for pairs with the separation $r_{\rm Nei}$\,$<$\,$30$\,$\textrm{kpc}\,h^{-1}$ with a weaker enhancement extending to $300$\,$\textrm{kpc}\,h^{-1}$ in late-type systems. 
\citet{2008MNRAS.385.1903L} found that nearly half of the most actively star-forming galaxies have companions within $100$\,$\textrm{kpc}\,h^{-1}$, often showing morphological signs of recent interactions. 
Similarly, \citet{2008AJ....135.1877E} reported SFR boosts of a factor of $\sim$\,2--3 for close ($r_{\rm Nei}$\,$<$\,$40$\,$\textrm{kpc}\,h^{-1}$) major and minor pairs. 
These enhancements appear robust across diverse environments and cosmic time (\citealt{1999ApJ...525...31S}; \citealt{2010MNRAS.407.1514E}; \citealt{2022ApJ...940....4S}).

However, proximity alone does not guarantee SF enhancement. 
Recent studies highlighted the critical role of the companion's gas content and star-forming activity. 
Gas-rich, late-type companions more effectively trigger SF of the target galaxies, while gas-poor, early-type companions often yield weak or negative responses (e.g., \citealt{2016ApJS..222...16C}; \citealt{2018ApJS..237....2Z})
For instance, \citet{2016ApJS..222...16C} found that spiral--spiral (S+S) pairs show significantly higher sSFR and SFE than spiral--elliptical (S+E) systems. 
\citet{2018ApJS..237....2Z} quantified this, reporting $\sim$\,4.6\,$\times$ higher mean SFE in S+S pairs. 
Following this, \citet{2019ApJ...882...14M} (hereafter \href{https://iopscience.iop.org/article/10.3847/1538-4357/ab3401}{Paper I}), the first paper in our \textit{``Living with Neighbors''} series \citep{2019ApJ...882...14M, 2019ApJ...887...59A, 2021ApJ...914...86A, 2021ApJ...909...34M, 2024ApJ...963..141Z}, confirmed that the enhancement of sSFR with decreasing separation in galaxy pairs is generally regulated by the SF activity of their companions, through both ram pressure (which enhances SF) and gas heating (which suppresses it).
Cosmological simulations supported this dichotomy: \citet{2023MNRAS.522.5107B} found $\sim$\,3\,$\times$ sSFR enhancement for star-forming companions, and a $\sim$\,12\,\% suppression for passive ones. 
Most recently, \citet{2024ApJ...963..141Z} (hereafter \href{https://iopscience.iop.org/article/10.3847/1538-4357/ad2063}{Paper V}) demonstrated that the hydrodynamical effects of pair interactions on sSFR enhancement vary with the relative orientation of the interacting galaxies’ spin axes, such that well-aligned systems exhibit higher sSFR enhancement.
These findings emphasize that both proximity and companion properties must be considered to fully understand interaction-induced SF.

Beyond fueling SF directly, interactions can induce structural transformation, most notably the formation of stellar bars. 
Tidal perturbations can destabilize disk galaxies and drive angular momentum redistribution, triggering bar formation (e.g., \citealt{1987MNRAS.228..635N}; \citealt{2004MNRAS.347..220B}; \citealt{2016ApJ...826..227L}; \citealt{2019MNRAS.483.2721P}). 
Bars are thought to form through $m=2$ (two-fold symmetric) disk instabilities induced by asymmetries in the gravitational potential during interactions. 
Observationally, \citet{2012MNRAS.423.1485S} reported elevated bar fractions in galaxies with close companions, particularly in bulge-dominated systems. 
Further studies found bars more common in clusters than in fields (e.g., \citealt{2012ApJ...761L...6M}; \citealt{2014ApJ...796...98L}; \citealt{2019NatAs...3..844Y}). 
However, other studies found no clear dependence on the environment and suggested that strong interactions may sometimes weaken or destroy bars (\citealt{2012ApJ...745..125L}; \citealt{2013MNRAS.429.1051C}), indicating a complex relationship between bar formation and interaction strength.

Once formed, bars significantly influence internal gas dynamics. 
Bars redistribute angular momentum and funnel cold gas toward galaxy centers, modulating central SF (e.g., \citealt{1993RPPh...56..173S}; \citealt{1999ApJ...525..691S}; \citealt{2004ApJ...600..595R}; \citealt{2012ApJ...747...60K}). 
Yet whether bars enhance or suppress SF remains debated. 
Numerous studies reported elevated central SF in barred galaxies, often showing higher central molecular gas densities and up to $\sim$0.5--1.0 dex higher SFRs than unbarred counterparts (\citealt{1995A&AS..111..115W}; \citealt{2012MNRAS.423.3486W}; \citealt{2019MNRAS.484.5192C}). 
The Mapping Nearby Galaxies at APO (MaNGA) data also identified ``turnover'' galaxies with a central upturn in H$\alpha$ and D$_n$(4000), indicating recent central starbursts (\citealt{2020MNRAS.499.1406L}). 
The bar length appears to be a key structural driver, with longer bars more strongly associated with SF enhancement (\citealt{2023ApJ...949...91Z}).

Conversely, other studies showed that bars are more common in red, gas-poor, and older stellar systems (e.g., \citealt{2016A&A...595A..63V}; \citealt{2020MNRAS.499.1116F}; \citealt{2021AJ....161..260Z}), in line with theoretical expectations that gas-rich disks resist bar formation due to higher disk stability (e.g., \citealt{2013MNRAS.429.1949A}; \citealt{2018A&A...609A..60K}). 
Moreover, recent work proposed that bars may actively quench SF. 
For example, \citet{2015A&A...580A.116G} coined the term ``bar quenching'' to describe strong bars in red, massive galaxies with suppressed SF. 
\citet{2021MNRAS.507.4389G} found that strong bars are more prevalent in quiescent galaxies, though this trend disappears when controlling for the bar length. 
\citet{2021A&A...651A.107G} showed that barred galaxies lie below the star-forming main sequence and often display suppressed central SF. 
Notably, \citet{2013ApJ...779..162C} suggested that bar impact is bimodal---enhancing SF in gas-rich, pseudo-bulge systems, and suppressing it in gas-poor, classical bulge systems.

Overall, the findings above imply that the effect of bars on SF is not universal but complex. 
The presence of a bar alone is insufficient to determine its impact during pair interactions; instead, the outcome depends on bar properties, gas content, and the companion's properties. 
In the context of galaxy interactions, this complexity is magnified---bars may amplify, suppress, or modulate SF depending on both internal dynamics and external gas supply. 
In this paper, we present a comprehensive study of how the presence and strength of stellar bars modulate SF activity in interacting systems. 
Focusing on the redshift range $0.02$\,$<$\,$z$\,$<$\,$0.08$, we examine whether bars enhance or suppress interaction-driven SF in galaxy pairs, and how this response varies with the star-forming properties of companion galaxies. 
Our analysis is based on a statistically robust sample of $\sim$\,300,000 galaxies imaged by the Dark Energy Camera Legacy Survey (DECaLS; \citealt{2019AJ....157..168D}).

The paper is organized as follows. 
In Section~\ref{sec:data}, we describe the dataset and methodology, including visual classifications from Galaxy Zoo DECaLS (GZD), bar identifications from deep learning models (\citealt{2022MNRAS.509.3966W}; \citealt{2023MNRAS.526.4768W}), and quantitative measurements of bar strength from isophotal analysis (\citealt{2025ApJ...982..129W}). 
Galaxy pairs are selected based on projected separation, relative line-of-sight velocity, and stellar mass ratio. The control sample of isolated galaxies is carefully constructed for comparison. 
In Section~\ref{sec:result}, we present our main findings, demonstrating that the impact of pair interactions on SF is strongly modulated by bar presence, bar strength, and companion properties. 
We show that bars play a dual role---either enhancing or suppressing SF---depending on the interaction context, with outcomes tightly coupled to the companion's star-forming activity. 
In Section~\ref{sec:face}, we discuss the physical implications and propose a new framework in which stellar bars act as dynamic mediators that connect internal disk structures on kpc-scales to external hydrodynamical influence operating over scales of hundreds of kpc-scales. 
In Section~\ref{sec:sum}, we summarize our main results. 
Throughout this paper, we adopt a standard $\Lambda$CDM cosmology with $\Omega_\textrm{m} = 0.3$, $\Omega_\Lambda = 0.7$, $H_{0} = 100 \, h \, \textrm{km} \, \textrm{s}^{-1} \textrm{Mpc}^{-1}$, and $h = 0.7$.

\section{Data and Methodology} \label{sec:data}
\subsection{Surveys and Bar Identification}

All data used in this study are drawn from the DESI Legacy Surveys (DESI-LS) Data Release 8 (DR8; \citealt{2019AJ....157..168D}), which combines three wide-field optical surveys---DECaLS, the Beijing-Arizona Sky Survey (BASS), and the Mayall z-band Legacy Survey (MzLS)---covering $\sim$\,14,000 deg$^2$ of the sky. 
BASS ($g$, $r$-bands) and MzLS ($z$-band) imaged the northern sky using the 2.3-m Bok and 4m Mayall telescopes, respectively.
The optical data are complemented by mid-infrared photometry (3.4$-$22 \textmu m) from NEOWISE, yielding deep imaging with typical $5\sigma$ point-source depths of $g = 24.0-24.3$ and $r = 23.4-23.8$ mag.
These depths are sufficient to detect both prominent and subtle stellar bars in disk galaxies.

Barred galaxies were identified using morphological classifications from Galaxy Zoo DESI (GZ DESI; \citealt{2023MNRAS.526.4768W}), which provides automated labels for 8.67 million galaxies from a Bayesian CNN trained on volunteer votes from Galaxy Zoo: DECaLS (\citealt{2022MNRAS.509.3966W}). 
The GZD labels follow the established Galaxy Zoo decision tree (\citealt{2008MNRAS.389.1179L}; \citealt{2011MNRAS.410..166L}) and include $\sim$314,000 visually classified galaxies. 
The CNN achieves $\sim$\,99\,\% accuracy relative to high-consensus volunteer labels and returns bar probabilities in three mutually exclusive classes: strong bar, weak bar, and no bar. 
Because bulge-dominated hosts can bias standard bar measures — 
central bulge light circularizes isophotes and depresses ellipse-fit bar ellipticities (\citealt{2011MNRAS.415.3308G}; \citealt{2012ApJ...746..136M}); torque-based strengths are diluted when a massive bulge increases the axisymmetric force, even in S0 galaxies (\citealt{2010ApJ...721..259B}); and in face-on views the thick inner bar component “barlens” can be confused with a classical bulge (\citealt{2015MNRAS.454.3843A}), complicating both decompositions and machine-learned labels — 
we restrict the sample to face-on disk galaxies using GZ DESI vote thresholds: $p_{\rm feature\text{-}or\text{-}disk}$\,$\geq$\,$0.27$ and $p_{\rm edge\text{-}on\_no}$\,$\geq$\,$0.68$, following \citet{2021MNRAS.507.4389G}.
This restriction minimizes contamination by bulge-dominated or edge-on systems in our analysis.
Bar classifications in GZ DESI are mutually exclusive and normalized ($p_{\mathrm{strong\_bar}}+ p_{\mathrm{weak\_bar}} + p_{\mathrm{bar\_no}} = 1$). 
We first select barred galaxies via $p_{\mathrm{bar\_no}}$\,$>$\,$0.5$ and then quantify bar strength using the conditional probability of strong vs. weak given that a system is barred:
For those galaxies, bar strength is defined as the conditional probability
\begin{equation} f_{\mathrm{strong}} \equiv \frac{p_{\mathrm{strong\_bar}}}{p_{\mathrm{strong\_bar}} + p_{\mathrm{weak\_bar}}} = P(\mathrm{strong}\,|\,\mathrm{barred})\,. \end{equation}
This avoids conflating bar presence with strength and reduces sensitivity to the absolute calibration of $p_{\mathrm{bar\_no}}$.
Galaxies with $f_{\mathrm{strong}}$\,$>$\,$0.5$ are labeled strong bars, and those with $f_{\mathrm{strong}}$\,$<$\,$0.5$ as weak bars.

Quantifying bar measurements are adopted from \citet{2025ApJ...982..129W}, which performed isophotal analyses on 232,142 DESI-LS DR8 galaxies. 
Bar parameters---length ($L_{\rm Bar}$), ellipticity ($e_{\rm Bar}$), and position angle---are derived via ellipse fitting using the IRAF {\fontfamily{qcr}\selectfont ellipse} task.
Bars are identified as local maxima in the radial ellipticity profile, typically accompanied by a plateau in position angle. 
Bar lengths are normalized by the host galaxy's optical diameter ($D_{25}$) and deprojected for inclination following \citet{2011ApJS..197...22L} and \citet{2023ApJ...949...91Z}.
The catalog includes uncertainties and quality flags, and shows strong agreement with GZ DESI classifications---93\,\% for bar presence and $>$\,90\,\% for bar strength. 
Stellar mass dependence of bar parameters is tested, given prior findings that normalization by mass or bulge-to-total ratio reduces scatter in bar-SF correlations (\citealt{2018MNRAS.474.5372E}; \citealt{2023ApJ...949...91Z}). 
We find no significant residual correlation between bar measurements and stellar mass in our sample, indicating that the catalog is suitable for unbiased analysis of bar-SF relationships.

Stellar masses are estimated using the empirical color-mass-to-light ($M/L$) relation from \citet{2003ApJS..149..289B}, calibrated for SDSS photometry but applied here to DESI Data Release 1 (DR1; \citealt{2025arXiv250314745D}) $g$ and $r$-band fluxes from DESI-LS imaging. 
The $r$-band $M/L$ ratio is calculated as: \begin{equation} \textrm{Log}(M_*/L_r) = -0.306 + 1.097 \times (g - r) \hspace{1mm}, \end{equation} where $(g-r)$ is corrected for Galactic extinction using $E(B-V)$ values from the {\fontfamily{qcr}\selectfont desi\_dr1.photometry}\footnote{\url{https://datalab.noirlab.edu/query.php?name=desi_dr1.photometry}} catalog, based on \citet{1998ApJ...500..525S} dust maps. 
Extinction coefficients $A_g = 3.214 \times E(B-V)$ and $A_r = 2.165 \times E(B-V)$ follow \citet{2024arXiv240905140Z}.
Fluxes in nanomaggies are converted to AB magnitudes via: \begin{equation} m = 22.5 - 2.5\hspace{1mm}\textrm{Log}(\rm Flux).\end{equation}
Absolute magnitudes in the $r$-band are computed using spectroscopic redshifts from {\fontfamily{qcr}\selectfont desi\_dr1.zpix}\footnote{\url{https://datalab.noirlab.edu/query.php?name=desi_dr1.zpix}} (with quality flag {\fontfamily{qcr}\selectfont zwarn}$ = 0$) and cosmological distances. 
Stellar mass is then derived as: \begin{equation} \textrm{Log} (M_*/M_\odot) = \textrm{Log} (M_*/L_r) - 0.4 \times (M_r - M_{r,\odot}) \hspace{1mm}, \end{equation} assuming $M_{r,\odot} = 4.67$. 
To align the DESI photometric system with SDSS, we apply a color correction: $(g-r)_\mathrm{SDSS} = (g-r)_\mathrm{DESI} - 0.02$ (\citealt{2019AJ....157..168D}; \citealt{2017PASP..129f4101Z}).

While value-added catalog such as \citet{2024A&A...691A.308S} and \citet{2014ApJS..210....3M} offer SED-based stellar masses or bulge-disk decompositions, their limited redshift range ($z \lesssim 0.06$) and partial sky coverage make them unsuitable for our full sample. 
We therefore uniformly adopt the \citet{2003ApJS..149..289B} method across the sample. 
Validation against SDSS DR16 for a subsample of $\sim$\,5,000 galaxies shows a median offset of $\Delta \textrm{Log}(M_*/M_\odot) \approx 0.06$ dex, confirming that the adopted method yields statistically robust stellar masses within a typical uncertainty of $\sim$\,0.1 dex.
To ensure mass completeness and reliability, our analysis is restricted to galaxies with $\textrm{Log}(M_*/M_{\odot}) \geq 8.75$.

Emission-line fluxes and stellar kinematics are obtained from the SDSS-based catalogs of \citet{2006MNRAS.366.1151S} and \citet{2013MNRAS.431.1383T}, which use the Gas AND Absorption Line Fitting (GANDALF) and the Penalized Pixel-Fitting (pPXF) algorithms (\citealt{2004PASP..116..138C}). 
These provide fluxes for H$\alpha$, H$\beta$, [O\textsc{iii}]$\lambda5007$, and [N\textsc{ii}]$\lambda6583$.
To exclude AGN contamination, we classify galaxies using the BPT diagram (\citealt{1981PASP...93....5B}) with demarcation criteria from \citet{2003MNRAS.346.1055K}, removing AGN and composite types. 
Interaction-driven SF enhancement and bar evolution are both expected to occur primarily within central $\sim$\,1$-$3 kpc regions (\citealt{2011MNRAS.412..591P}; \citealt{2020ApJ...893...19W}).
We therefore quantify central SF using fiber-based SFRs and sSFRs from the MPA/JHU value-added catalog (\citealt{2004MNRAS.351.1151B}), derived from extinction-corrected H$\alpha$ luminosities within the 3\arcsec\ SDSS fiber aperture---corresponding to $\sim$\,1$-$3 kpc at $0.02 < z < 0.08$.

\subsection{Pair Identification and Control Sample}

We identify galaxy pairs following the methodology established in our previous studies (\href{https://iopscience.iop.org/article/10.3847/1538-4357/ab3401}{Paper I}; \href{https://iopscience.iop.org/article/10.3847/1538-4357/ad2063}{Paper V}).
For each galaxy, neighboring companions are identified based on radial velocity differences $\left\vert \Delta v \right\vert < 1000\,\rm km\,s^{-1}$ and a minimum stellar mass ratio of 1:10. 
Galaxies with no such companions within a projected separation of $r_{\rm Nei} \leq 250\,\textrm{kpc}\,h^{-1}$ are classified as \textit{purely isolated} and serve as the control sample. 
We define a \textit{purely isolated pair} using the following criteria: 
\begin{enumerate}
    \item 
    \textbf{Projected Separation:} The closest companion lies within $r_{\rm Nei} \leq 250\,\textrm{kpc}\,h^{-1}$, with a minimum angular separation of 3\arcsec\ to avoid photometric blending. 
    \item 
    \textbf{Radial Velocity Offset:} The line-of-sight velocity difference satisfies $\left\vert \Delta v \right\vert \leq 300\,\rm km\,s^{-1}$, ensuring a physically bound or pre-coalescence system. 
    \item 
    \textbf{Stellar Mass Ratio:} Pairs are limited to 1:10--10:1 mass ratios to focus on major interactions and avoid minor-merger biases. 
    \item 
    \textbf{Multiplicity Exclusion:} Systems with multiple qualifying neighbors are excluded to isolate the effects of one-to-one interactions. 
    \item 
    \textbf{Survey Edge Cut:} To avoid edge effects, only galaxies located at least $4 \rm Mpc \hspace{1mm} h^{-1}$ from the survey footprint boundary are included. 
\end{enumerate}
These criteria yield a clean sample of isolated galaxy pairs, minimally influenced by group environments or higher-order systems. 

\begin{figure}
	\includegraphics[width=\columnwidth]{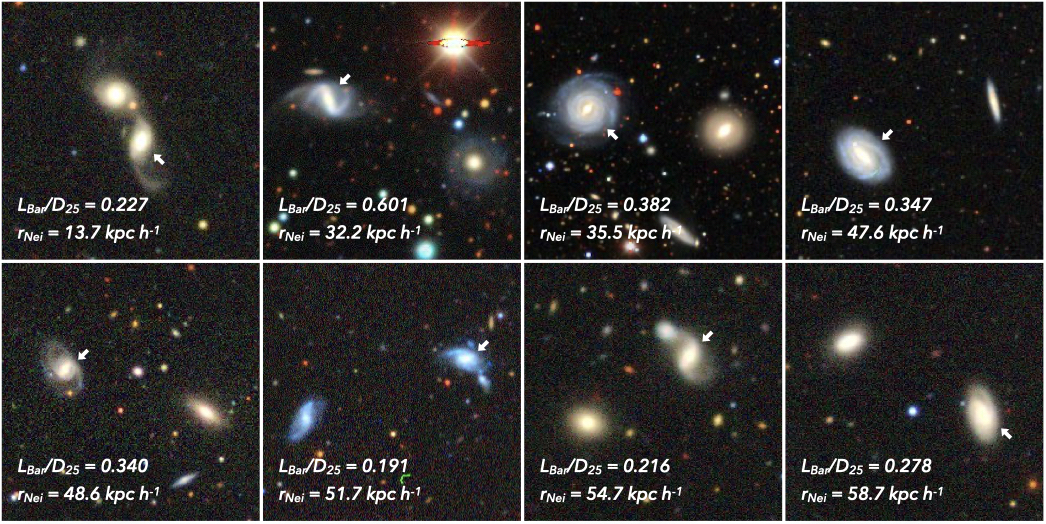}
    \caption{The example DESI-LS multiband images of galaxy pairs in which the target galaxy host a stellar bar. 
    For each system, we indicate the normalized bar length ($L_{\rm Bar}/D_{25}$) and the projected separation between the target and its companion ($r_{\rm Nei}$). 
    The target barred galaxy is marked with a white arrow in each panel.}
    \label{fig:1}
\end{figure}

Figure~\ref{fig:1} shows example DESI image cutouts ($g$, $r$, $z$-bands) for barred galaxies in close pairs ($10\,\textrm{kpc}\,h^{-1} < r_{\rm Nei} < 60\,\textrm{kpc}\,h^{-1}$), with $L_\textrm{Bar}/D_\textrm{25}$ labeled and barred galaxies indicated with white arrows.
Among 23,807 purely isolated galaxies, 8,023 ($\sim$33.7\,\%) are barred.
In comparison, 1,572 of 4,064 galaxies in pairs ($\sim$38.7\,\%) host a bar, indicating a modest increase in bar fraction due to pair interaction.
However, dividing by bar strength reveals more significant trends: 494 paired galaxies ($\sim$12.2\,\%) host strong bars and 1,078 ($\sim$26.5\,\%) host weak bars, compared to 1,507 strong ($\sim$6.3\,\%) and 6,515 weak bars ($\sim$27.4\,\%) among isolated galaxies.
This suggests that while the overall bar fraction remains similar, interactions preferentially enhance the incidence of strong bars.
However, these raw fractions do not account for differences in intrinsic properties; we therefore perform a matched-control analysis later in this section.

The primary objective of this study is to disentangle the individual and combined effects of galaxy-galaxy interactions and bar structures on SF in galaxy pairs, and to examine how these effects vary with companion properties. 
To achieve this, we construct a carefully matched control sample of isolated, unbarred galaxies, following the methodology of \citet{2019ApJ...882...14M} and \citet{2024ApJ...963..141Z}. 
Control galaxies are selected to match the pair sample in redshift, stellar mass, and local environment, minimizing systematic biases. 
The control sample is drawn from galaxies classified as both isolated and unbarred, defined as having no companions within $r_{\rm Nei} \leq 250 \,\textrm{kpc}\,h^{-1}$ and $\left\vert \Delta v \right\vert < 1000\,\textrm{km s}^{-1}$, and a GZ DESI vote fraction of $p_{\rm bar_{no}} > 0.5$. 
For each galaxy in the pair sample---regardless of bar classification---we randomly select a control galaxy within $\pm 0.005$ in redshift, $\pm 0.1$ dex in stellar mass, and $\pm 0.1$ dex in local galaxy density. 
The local density is calculated using adaptive kernal density estimation (\citealt{1986desd.book.....S}), providing a continuous, non-parametric measure of environment across a broad dynamic range.

\begin{figure}
	\includegraphics[width=\columnwidth]{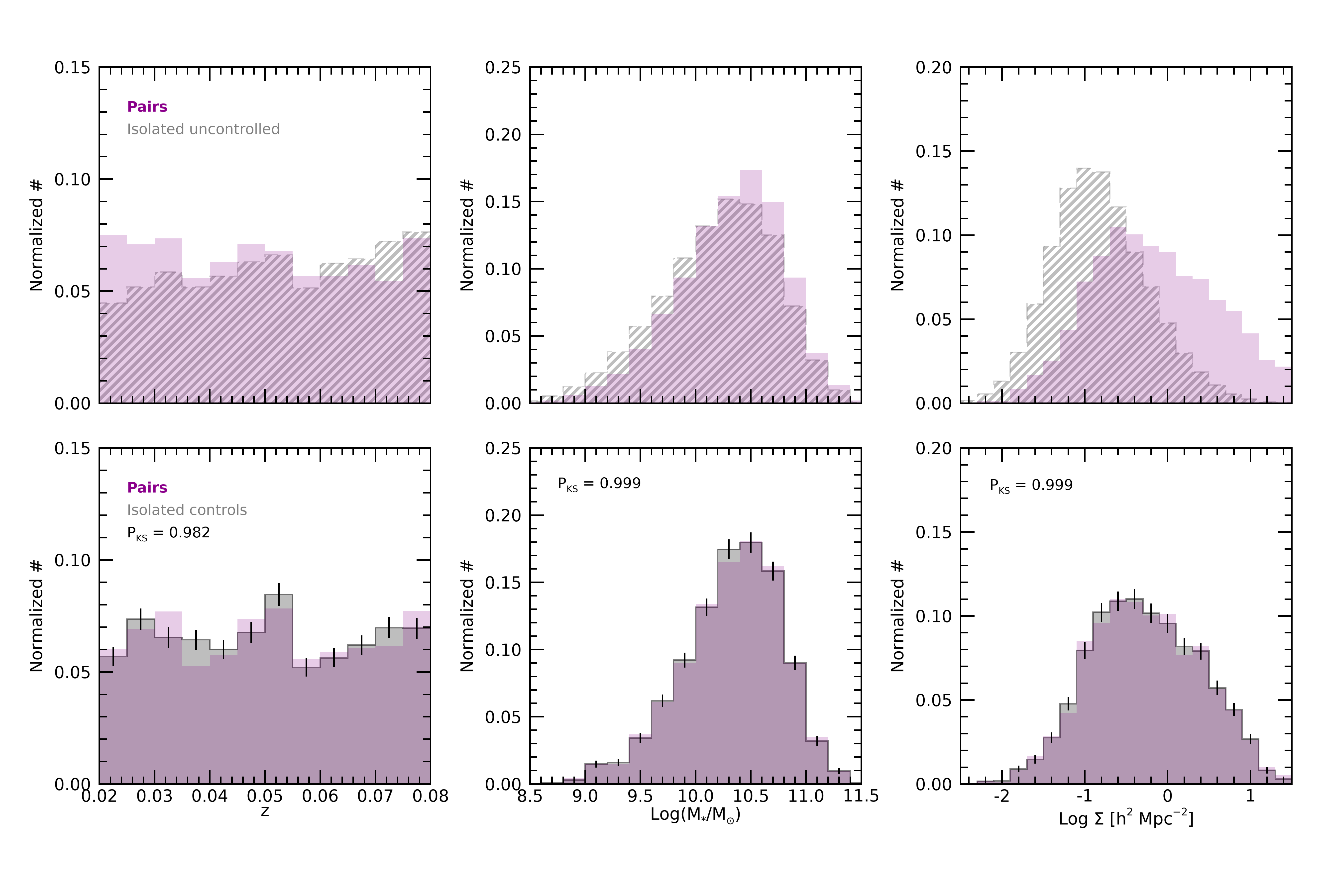}
    \caption{Top: The normalized distributions of redshift (left), stellar mass (middle), and local density (right) for paired galaxies (purple) and the initial, unmatched sample of isolated, unbarred galaxies (dashed gray). 
    Bottom: Same as the top panels, but comparing the matched sample of paired galaxies (purple) to the final control sample of isolated, unbarred galaxies (solid gray). 
    The $p$-values from K-S test confirm the statistical consistency between the matched samples, validating the effectiveness of our control selection. 
    Black error bars on the histograms represent Poisson uncertainties for the control sample.}
    \label{fig:2}
\end{figure}

Figure~\ref{fig:2} compares redshift, stellar mass, and density distributions for different samples. 
The top panels are for the uncontrolled isolated/unbarred sample and the entire pair sample.
As expected, paired galaxies are skewed toward lower redshifts, higher masses, and denser environments, consistent with known observational biases---e.g., pair detectability and bar identification are limited at higher redshifts (\citealt{2020MNRAS.491.2481B}; \citealt{2020A&A...634A.123R}; \citealt{2024A&A...688A.158L}), and bar fraction increases with stellar mass (\citealt{2014MNRAS.438.2882M}; \citealt{2018MNRAS.474.5372E}). 
The bottom panels show that after matching, the distributions for the pair and control samples become statistically indistinguishable, ensuring fair comparison. 
In total, we identify 3,893 galaxies in isolated pairs—comprising 475 strongly and 1,032 weakly barred galaxies—that are successfully matched to isolated, unbarred counterparts. The final control sample comprises 389,300 galaxies, generated through 100 iterations of one-to-one random matching.
In a similar way, we constructed an additional sub-control sample that included both barred and unbarred galaxies, considering only whether a system was paired or isolated, for use in a complementary test.
This allows a direct assessment of how bar presence and strength depend on pair membership. 
Among 389,300 isolated controls, 130,173 ($\sim$33.4\,\%) are barred, comprising 29,939 ($\sim$7.7\,\%) strong and 100,234 ($\sim$25.7\,\%) weak bars. 
Compared against the pair sample introduced above, the overall barred fraction remains comparable, whereas the enhancement of strong bars in paired systems persists after matching in stellar mass, redshift, and local density.

\section{Result}
\label{sec:result}

\begin{figure}
	\includegraphics[width=\columnwidth]{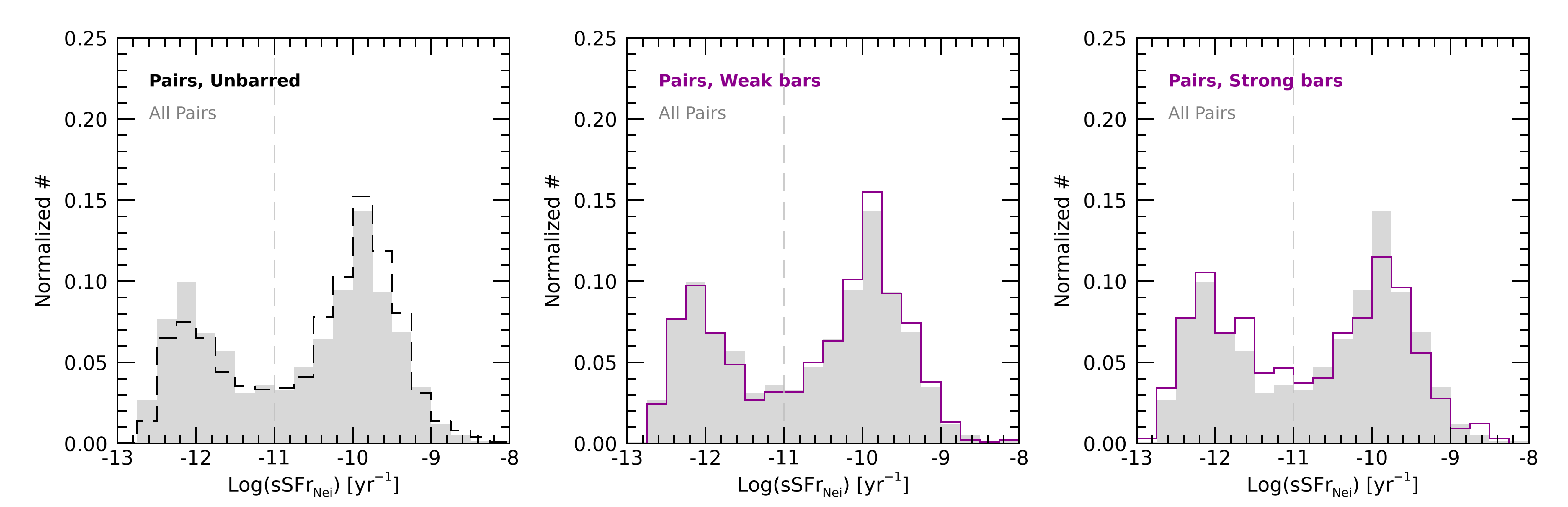}
    \caption{The normalized distributions of the sSFR of the nearest companion galaxy for paired systems, categorized by bar classification of the target galaxy: unbarred (left; dashed black), weakly barred (middle; solid purple), and strongly barred (right; solid purple). 
    For reference, the distribution for all paired systems, irrespective of bar classification, is shown as solid gray histogram in each panel. 
    The vertical gray dashed line indicates the sSFR threshold, $\textrm{Log}(\rm sSFR_{\rm Nei}) = -11 \hspace{1mm} \rm yr^{-1}$, used to separate star-forming and quiescent companions in our analysis.}
    \label{fig:3}
\end{figure}

Building on previous studies that emphasized the role of companion SF activity in modulating interaction-driven responses (e.g., \citealt{2007AJ....134..527W}; \citealt{2009ApJ...691.1828P}; \citealt{2010ApJ...713..330X}; \citealt{2011MNRAS.412..591P}; \citealt{2016ApJS..222...16C}; \citealt{2018ApJS..237....2Z}; \href{https://iopscience.iop.org/article/10.3847/1538-4357/ab3401}{Paper I}; \href{https://iopscience.iop.org/article/10.3847/1538-4357/ad2063}{Paper V}), we extend this investigation by examining how the influence of companion properties varies with the presence and strength of stellar bars. 
To this end, we divide our sample based on the sSFR of the nearest neighbor ($\rm sSFR_{\rm Nei}$).
Figure~\ref{fig:3} shows the $\rm sSFR_{\rm Nei}$ distribution for unbarred, weakly barred, and strongly barred galaxies. 
In all cases, the distributions are bimodal, with a minimum near $\textrm{Log}(\rm sSFR_{\rm Nei}) \approx -11 \hspace{1mm} \rm yr^{-1}$, which we adopt as the threshold to classify neighbors as either star-forming or quenched (vertical dashed lines in the figure). 
A systematic trend is evident across bar classes: the fraction of passive neighbors increases monotonically from unbarred ($\sim33.3$\,\%) to weakly barred ($\sim$\,37.4\,\%) to strongly barred galaxies ($\sim$\,45.6\,\%).
While the bimodal structure is preserved in all subsamples, the rising incidence of quenched companions with increasing bar strength suggests a potential link between bar formation and interaction environments.

\subsection{Interaction-Induced SF Depending on Bar Classes}

\begin{figure}
	\includegraphics[width=\columnwidth]{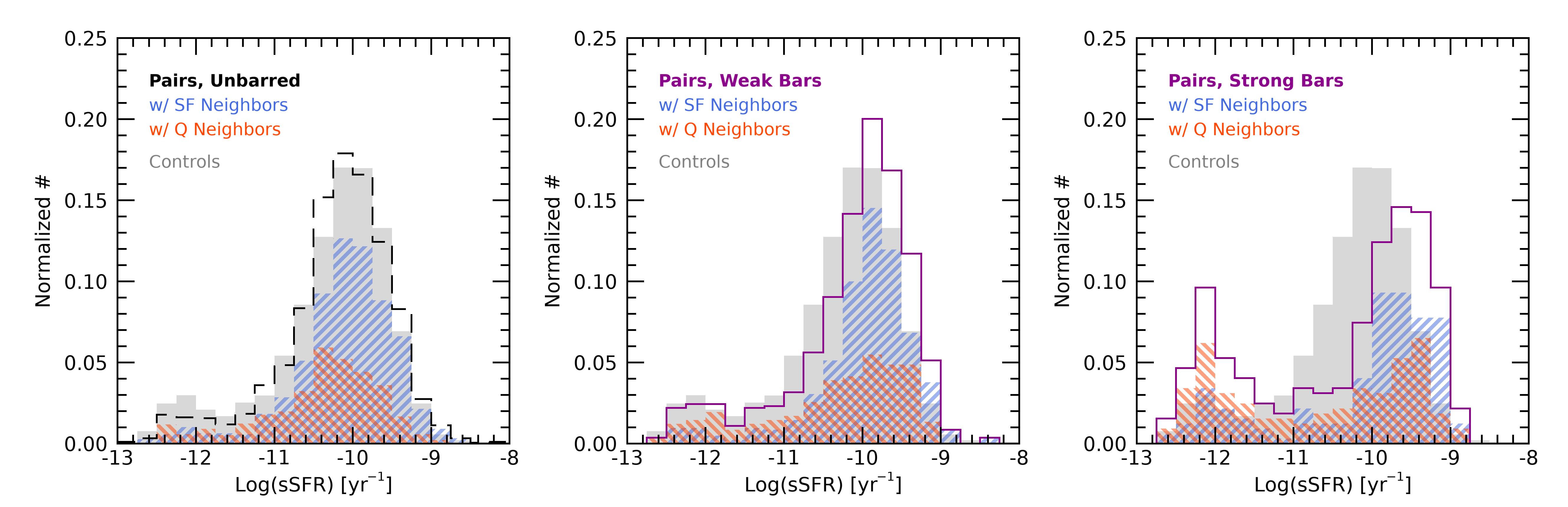}
    \caption{The normalized distributions of the sSFR of target galaxies in pairs, categorized by bar classification: unbarred (left; dashed black), weakly barred (middle; solid purple), and strongly barred (right; solid purple). 
    For comparison, the corresponding distribution for the control sample of isolated, unbarred galaxies is shown as solid gray histogram in each panel. 
    Paired galaxies are further subdivided by the SF activity of their companions: systems with star-forming companions are shown in dashed blue, while those with quiescent companions are shown in dashed red.}
    \label{fig:4}
\end{figure}

Figure~\ref{fig:4} shows the distribution of sSFR for galaxies in pairs, categorized by the bar strength. 
Due to selection biases favoring nearly face-on disks, our sample is biased toward systems with higher sSFR, typically above the main sequence threshold at $\textrm{Log}(\rm sSFR_{\rm}) \approx -11 \hspace{1mm} \rm yr^{-1}$.
Nevertheless, a clear trend emerges: as bar strength increases, the sSFR distribution becomes more dispersed and distinctly bimodal. 
Consistent with the results of \href{https://iopscience.iop.org/article/10.3847/1538-4357/ab3401}{Paper I}, unbarred galaxies in pairs resemble their isolated, unbarred counterparts, while barred galaxies exhibit both a shift toward higher sSFR and a larger population at lower sSFR, thereby increasing the bimodality.
Using the same $\textrm{Log}(\rm sSFR_{\rm Nei}) = -11\,\rm yr^{-1}$ threshold, we divide galaxies into two subsamples having star-forming and passive neighbors. 
The passive fraction rises systematically from $\sim12.0$\% in unbarred, to $\sim13.1$\% in weakly barred, and $\sim30.5$\% in strongly barred galaxies. 
Among the targets with star-forming companions, the mean sSFR also increases with bar strength, reaching $\textrm{Log}(\rm sSFR) \approx -10.0$, $-9.86$, and $-9.64\,\rm yr^{-1}$ for unbarred, weakly barred, and strongly barred galaxies, respectively.

\begin{figure}
	\includegraphics[width=\columnwidth]{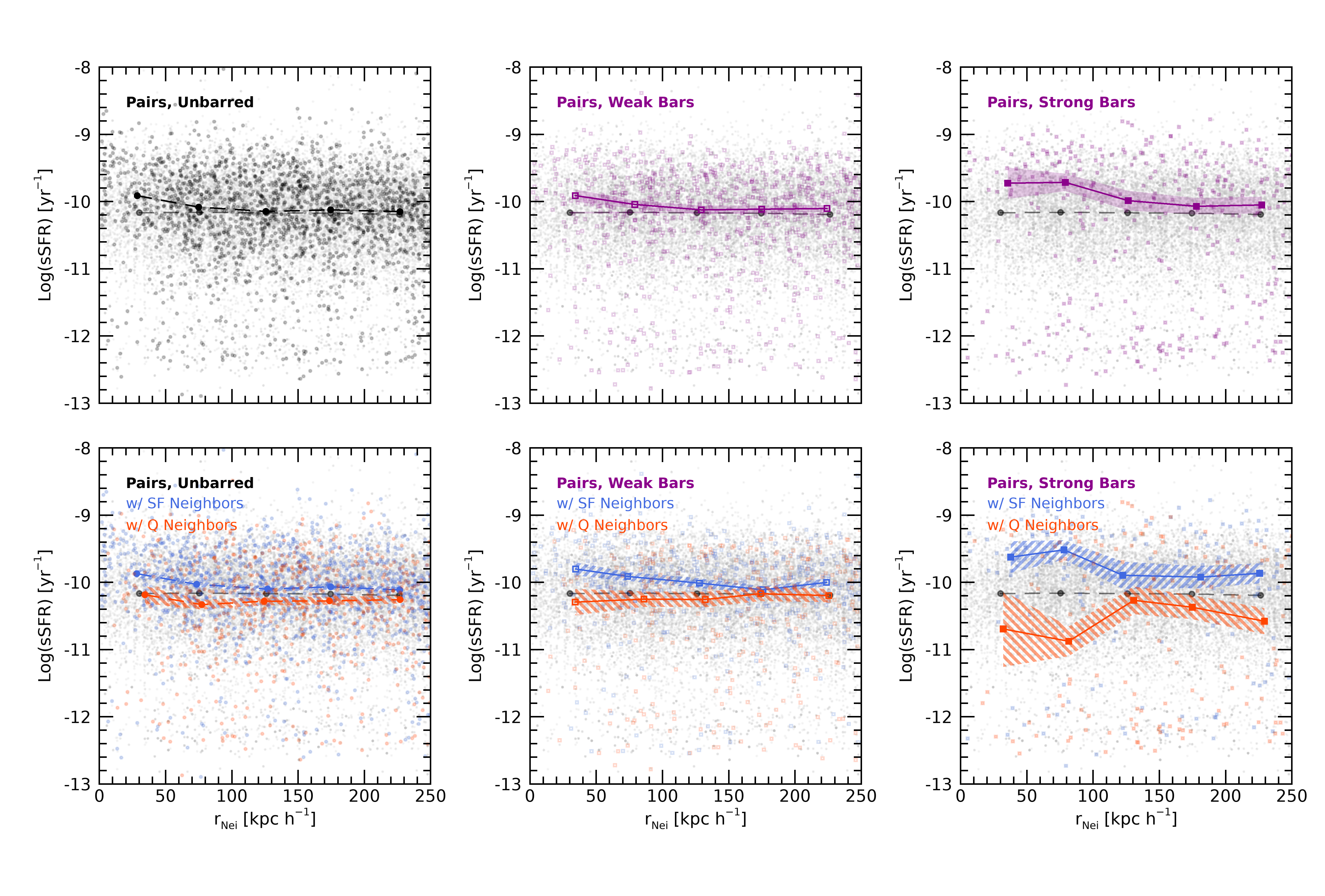}
    \caption{Top: The distributions of projected separation vs. the sSFR of target galaxies in pairs, categorized by bar classification: unbarred (left; black dots), weakly barred (middle; purple dots), and strongly barred (right; purple dots). 
    Dashed black and solid purple lines indicate the median trends for each subsample. 
    For comparison, isolated, unbarred control sample are shown as gray points with dashed gray median lines. 
    Shaded bands represent the statistical uncertainty derived from jackknife resampling. 
    Bottom: The same as the top panels, but the pair sample is further subdivided based on the SF activity of the companion galaxy: systems with star-forming companions are shown in blue, and those with quiescent companion in red. 
    Median trends for each subsample are shown as blue and red lines, respectively.}
    \label{fig:5}
\end{figure}

\begin{figure}
	\includegraphics[width=\columnwidth]{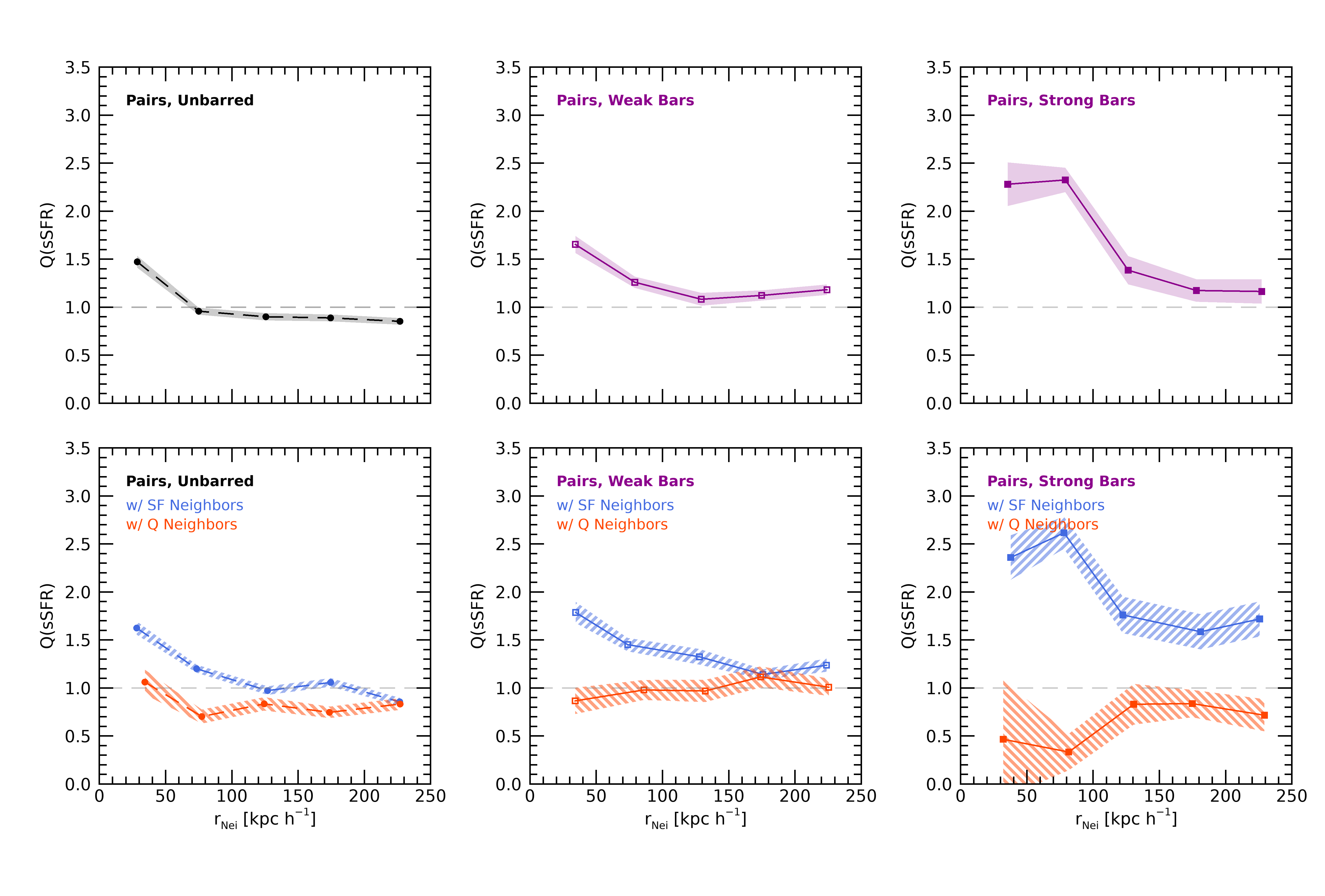}
    \caption{Top: The interaction-induced SF enhancement, $Q(\rm sSFR)$---defiend as the ratio between the sSFR of the target galaxy and that of its matched control sample---plotted as a function of projected separation from the nearest companion, categorized by bar classification: unbarred (left; dashed black), weakly barred (middle; solid purple), and strongly barred (right; solid purple). 
    Shaded bands represent the statistical uncertainty derived from jackknife resampling. 
    The horizontal dashed gray line at $Q(\rm sSFR) = 1$ indicates the baseline level where no enhancement is observed.
    Bottom: The same as the top panels, but the pair sample is further subdivided based on the SF activity of the companion galaxy: systems with star-forming companions are shown in blue, and those with quiescent companion in red. }
    \label{fig:6}
\end{figure}

Figure~\ref{fig:5} presents the sSFR of paired galaxies as a function of projected separation, compared to a matched control sample of isolated, unbarred galaxies. 
The upper panels show that, while the control sample has no trend with separation, paired galaxies---particularly those with stronger bars---exhibit increasing sSFR at smaller separations. 
The lower panels display that companion properties further modulate the effect: galaxies with star-forming companions show pronounced sSFR enhancement with decreasing separation, whereas those with passive companions show little to no enhancement or even suppression. 
To qualify these effects, we define a relative enhancement index $Q({\rm sSFR}) \equiv {\rm sSFR}/{\rm sSFR}_{\rm Control}$ and show it as a function of separation in Figure~\ref{fig:6}.
For unbarred and weakly barred galaxies with star-forming companions, close interactions ($r_{\rm Nei} < 125 \,\textrm{kpc}\,h^{-1}$) yield modest enhancement ($Q({\rm sSFR}) \sim 1.5$), while those with passive companions remain flat or slightly suppressed. 
In strongly barred galaxies, the divergence is remarkable: $Q({\rm sSFR})$ reaches $\sim2.5$ with star-forming companions but drops below $\sim0.5$ with passive ones. 
Despite larger scatter in the strongly barred subsamples, the results reveal two distinct pathways for barred galaxies in pairs---either triggering or suppressing SF---depending on the star-forming activity of the companion.

\subsection{No Direct Correlation between Bars and Pair Interactions}

\begin{figure}
	\includegraphics[width=\columnwidth]{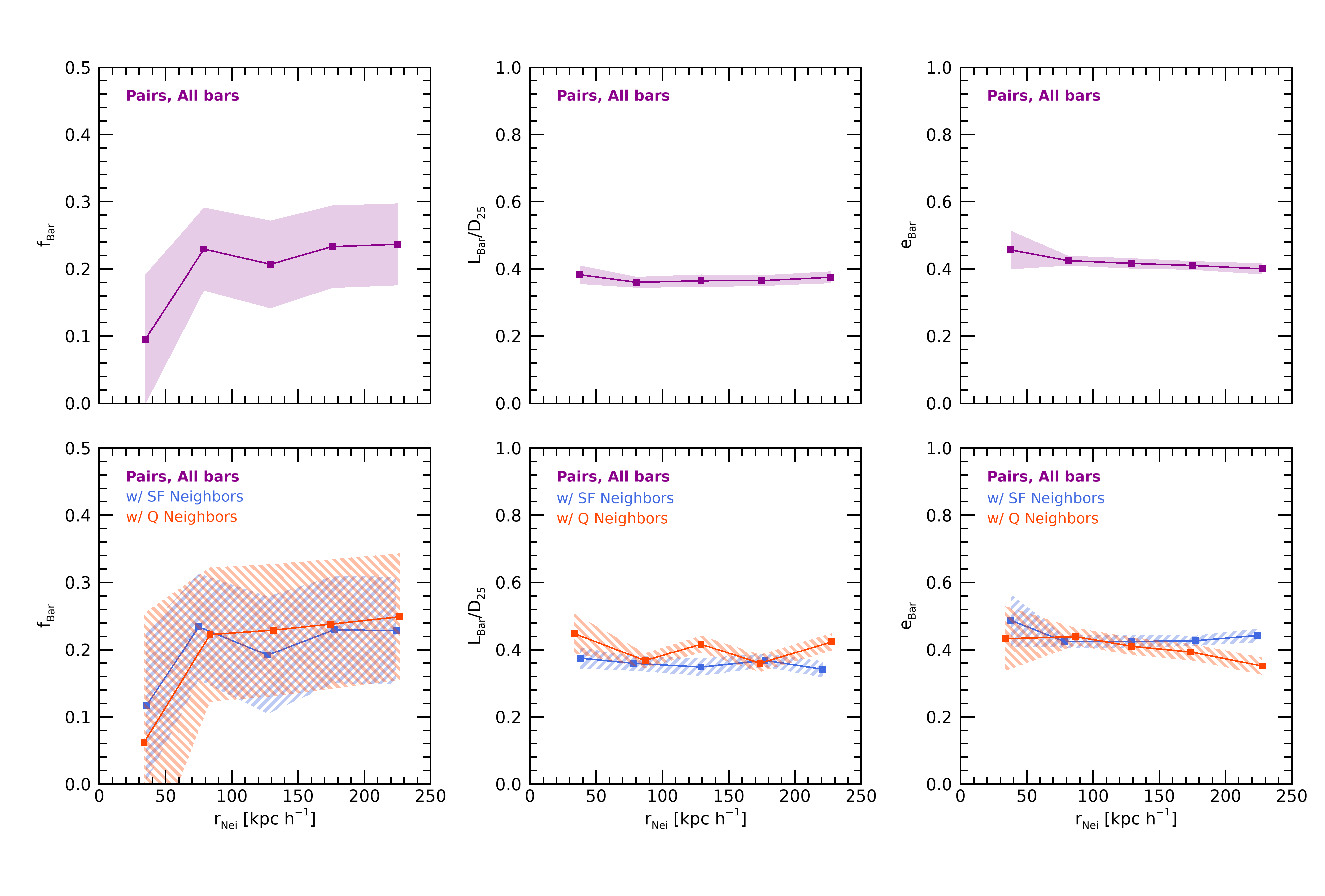}
    \caption{Top: The distributions of bar fraction ($f_{\mathrm{Bar}}$; left), normalized bar length ($L_{\rm Bar}/D_{25}$; middle), and bar ellipticity ($e_{\mathrm{Bar}}$; right) as a function of projected separation from the nearest companion galaxy. 
    Shaded bands represent statistical uncertainties, derived from Poisson errors for the bar fraction (left) and from jackknife resampleing for the structural parameters (middle and right).
    Bottom: The same as the top panels, but the pair sample is further subdivided based on the SF activity of the companion galaxy: systems with star-forming companions are shown in blue, and those with quiescent companion in red.}
    \label{fig:7}
\end{figure}

Bar formation is conventionally attributed to tidal perturbations arising from major mergers or satellite flybys (e.g., \citealt{1990A&A...230...37G}; \citealt{2004MNRAS.347..220B}; \citealt{2016ApJ...826..227L}; \citealt{2017MNRAS.464.1502M}; \citealt{2018ApJ...857....6L}; \citealt{2019MNRAS.483.2721P}; \citealt{2019A&A...624A..37L}; \citealt{2025ApJ...978...37A}).
For instance, \citet{2014ApJ...790L..33L} used collisionless N-body simulations to show that bar growth can be induced by flyby interactions across a range of mass ratios. 
Observationally, \citet{2019NatAs...3..844Y} found elevated bar fractions in interacting clusters, suggesting a role for large-scale gravitational influences. 
Such findings imply that interaction-driven and bar-regulated SF may coexist in galaxy pairs. 
To distinguish these effects, we examine the dependence of bar properties on projected separation---an observational proxy for the tidal strength. 
As shown in Figure~\ref{fig:7}, we find no significant correlation between bar fraction and separation. 
In fact, regardless of neighbors' SF properties, the bar fraction decreases at $r_{\rm Nei} < 50\,\textrm{kpc}\,h^{-1}$, contrary to expectations from tidal triggering scenarios. 
Moreover, the key bar structural parameters---the normalized bar strength ($L_{\rm Bar}/D_{25}$) and ellipticity ($e_{\rm Bar}$)---remain flat with separation and show no dependence on the companion's SF properties.

The results are consistent with recent studies suggesting that close companions may suppress, rather than induce, bars. 
\citet{2013MNRAS.429.1051C} found that bar detection decreases in very close pairs, while spiral features become more prominent. 
\citet{2024PhyS...99f5014T} further reported that no bar excess in paired galaxies, and even a deficit in weak bars at fixed mass. 
\citet{2023A&A...679A...5A} observed that bar sizes are larger in field galaxies, though the difference vanishes when normalized by galaxy size. 
On the other hand, simulations offered additional support. 
\citet{2018MNRAS.474.5645P} showed that flyby encounters do not systematically alter bar length or pattern speed. 
From the TNG100 simulation, \citet{2025arXiv250402145C} concluded that bar fractions in high surface brightness disks are largely unaffected by tidal forces. 
Similarly, \citet{2024MNRAS.529..979L} and \citet{2022MNRAS.510.5164C} found no consistent link between bar formation and interaction strength, and major mergers between galaxies can, in some cases, lead to the destruction of bars.
The findings suggest that bar formation and strength are governed primarily by galaxies' internal properties---such as disk stability, bulge-to-total ratio, and gas content---rather than by external perturbations. 
Our results align with this view, indicating that ongoing interactions did not trigger bars in our sample.
This decoupling between bar presence and tidal influence strengthens the reliability of our analysis on bar-regulated SF and its dual role in either enhancing or suppressing SF in paired galaxies.

\subsection{SF Regulation by Bars in Paired Galaxies}

\begin{figure}
	\includegraphics[width=\columnwidth]{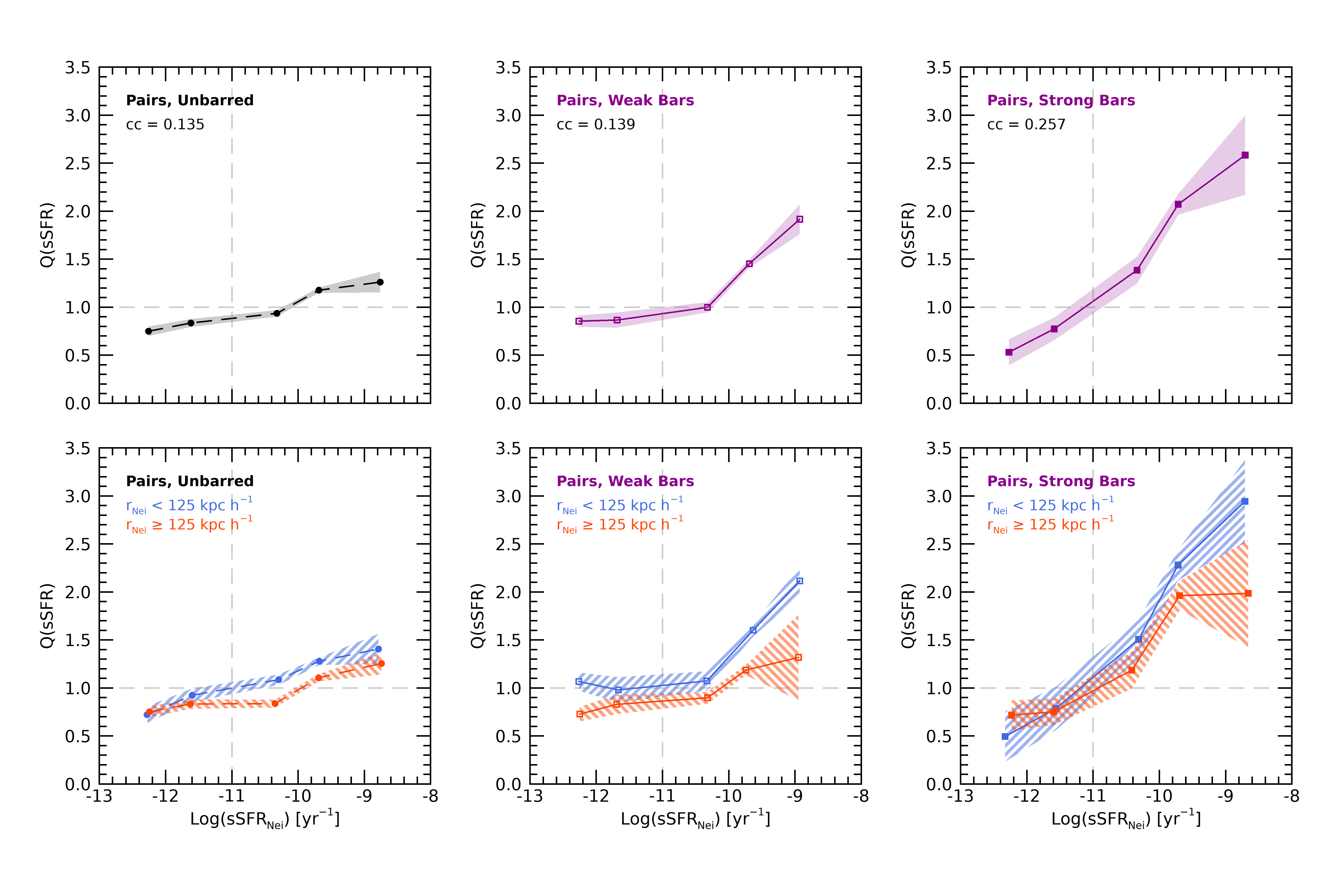}
    \caption{Top: The distributions of interaction-induced SF enhancement, $Q(\rm sSFR)$, as a function of the sSFR of the nearest companion, categorized by bar classification: unbarred (left; dashed black), weakly barred (middle; solid purple), and strongly barred (right; solid purple). 
    Shaded bands represent the statistical uncertainty derived from jackknife resampling.
    The Pearson coefficient for each subsample is indicated in the corresponding panel. 
    The horizontal dashed gray line at $Q(\rm sSFR) = 1$ indicates the baseline level with no enhancement, while the vertical dashed line at $\textrm{Log}(\rm sSFR_{\rm Nei}) = -11 \hspace{1mm} \rm yr^{-1}$ denotes the threshold used to classify companions as star-forming or quiescent.
    Bottom: The same as the top panels, but the pair sample is further divided based on the projected separation from the nearest companion galaxy: close pairs ($r_{\rm Nei}$\,$<$\,$125\,\textrm{kpc}\,h^{-1}$) are shown in blue, and wide pairs ($r_{\rm Nei}$\,$\geq$\,$125\,\textrm{kpc}\,h^{-1}$) in red.}
    \label{fig:8}
\end{figure}

Figure~\ref{fig:8} presents $Q(\rm sSFR)$ as a function of $\rm sSFr_{\rm Nei}$ to examine the joint influence of bar strength and companion SF activity on interaction-driven SF.
The upper panels show that, across all bar classes---unbarred, weakly barred, and strongly barred---$Q(\rm sSFR)$ increases with $\rm sSFr_{\rm Nei}$, reaffirming \href{https://iopscience.iop.org/article/10.3847/1538-4357/ab3401}{Paper I} in that actively star-forming companions drive enhanced SF. 
However, this trend steepens with increasing bar strength. 
Compared to unbarred and weakly barred galaxies (left and middle panels, respectively), strongly barred galaxies (right panel) exhibit $Q(\rm sSFR)$ that rises much more steeply with the neighbors' sSFR.
The increasing Pearson coefficients---$cc = 0.135$, $0.139$, and $0.257$ for unbarred, weakly barred, and strongly barred galaxies, respectively---suggest that bars amplify a galaxy's responsiveness to actively star-forming companions.
In the lower panels, to assess the role of proximity, we divide the sample into close ($r_{\rm Nei} \leq 125\,\textrm{kpc}\,h^{-1}$) and wide pairs.
In all bar categories, close pairs show stronger correlations, reinforcing that tidal effects are more pronounced at smaller separations. 
This indicates that the efficiency of interaction-induced SF depends jointly on companion SF activity, bar strength, and pair separation---highlighting a dynamic interplay between internal structures and external influences.

\begin{figure}
	\includegraphics[width=\columnwidth]{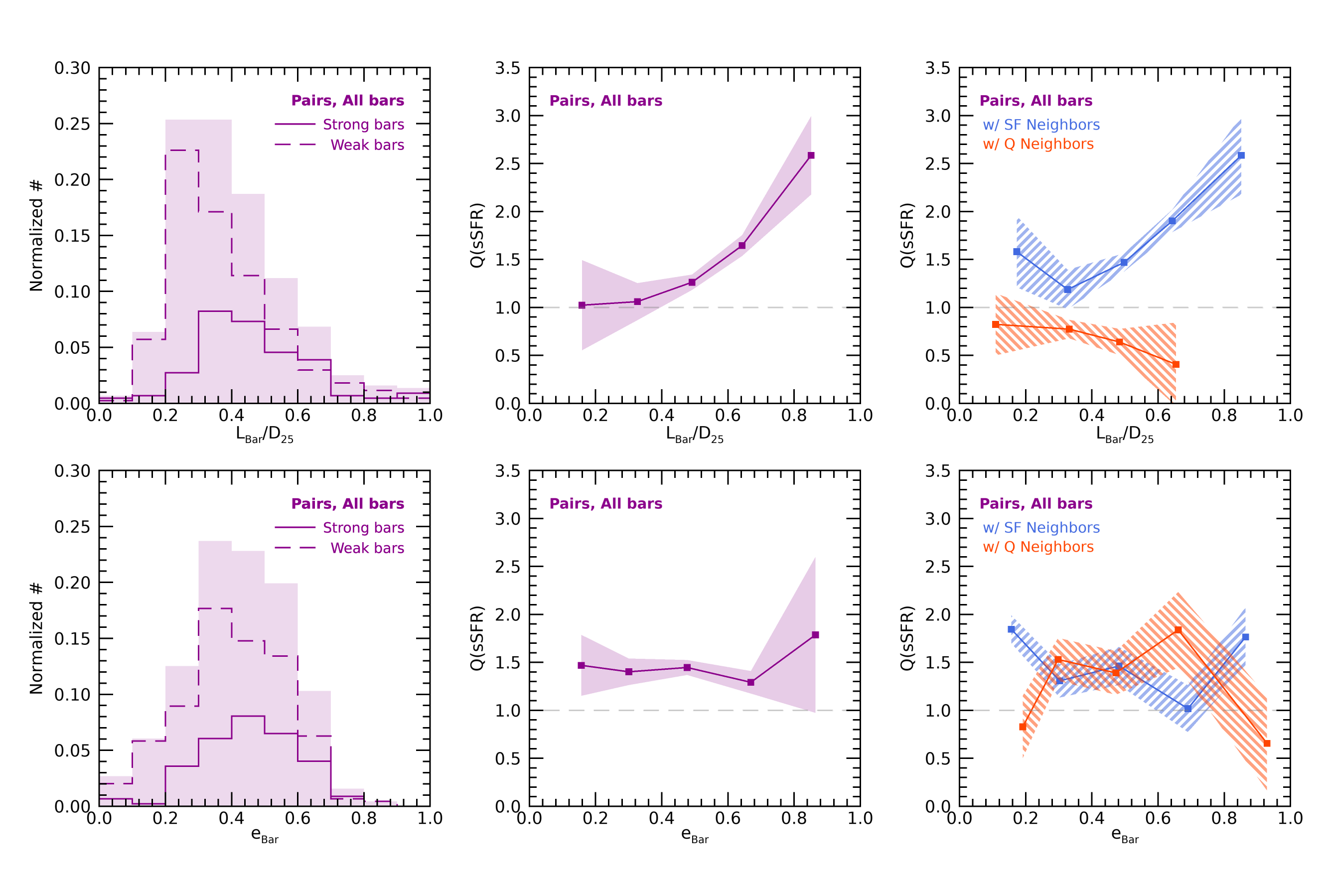}
    \caption{The distributions of bar structural parameters and the corresponding interaction-induced SF enhancement, $Q(\rm sSFR)$, as a function of those parameters.
    Top and bottom rows show results for normalized bar length ($L_{\rm Bar}/D_{25}$) and bar ellipticity ($e_{\mathrm{Bar}}$), respectively. 
    Left: The normalized histograms of bar properties for all paired barred galaxies (shaded purple), with dashed and solid purple histograms indicating weakly and strongly barred subsample, respectively, as classified by \citet{2023MNRAS.526.4768W}.
    Middle: $Q(\rm sSFR)$ as a function of each bar property. 
    Shaded bands represent the statistical uncertainty derived from jackknife resampling. 
    The horizontal dashed gray line at $Q(\rm sSFR) = 1$ indicates the baseline level with no enhancement. 
    Right: The same as the middle panels, but the pair sample is subdivided by the SF activity of the companion galaxy: systems with star-forming companions are shown in blue, and those with quiescent companion in red.}
    \label{fig:9}
\end{figure}

Building on GZ DESI bar classifications (\citealt{2023MNRAS.526.4768W}), we further investigate how $Q(\rm sSFR)$ correlates with physical bar properties using measurements from \citet{2025ApJ...982..129W}. 
The left panels of Figure~\ref{fig:9} show that visually strong bars correspond to longer normalized bar lengths ($L_{\rm Bar}/D_{25}$), while ellipticity ($e_{\rm Bar}$) distributions show little distinction---consistent with previous work that identified bar length as a more robust strength indicator than ellipticity (\citealt{2005MNRAS.364..283E}; \citealt{2009A&A...495..491A}; \citealt{2011MNRAS.415.3308G}; \citealt{2016A&A...587A.160D}; \citealt{2023ApJ...949...91Z}). 
In the middle panels, we find that $Q(\rm sSFR)$ increases with $L_{\rm Bar}/D_{25}$, reaching $\sim2.5$ for the longest bars, while remaining flat with $e_{\rm Bar}$, further underscoring the bar length as the key structural driver. 
In the right panels, the distinction is magnified when separating galaxies by companion activity: for star-forming companions, $Q(\rm sSFR)$ rises steeply with bar length, while for passive companions, $Q(\rm sSFR)$ shows a modest decline, suggesting possible bar-driven suppression.

Taken together, our findings indicate that bar length, not ellipticity, is the principal structural parameter that modulates interaction-driven SF. 
The bars appear to play a dual role: amplifying SF in gas-rich interactions while potentially suppressing SF in gas-poor interactions. 
The results underscore the importance of jointly considering both internal bar structures and external companion properties in order to understand the SF regulation in galaxy pairs.

\section{Discussion: Dual Role of Bars in Paired Galaxies}\label{sec:face}

Our previous study (\href{https://iopscience.iop.org/article/10.3847/1538-4357/ab3401}{Paper I}) demonstrated that SF enhancement in paired galaxies is not only regulated by proximity but is also strongly influenced by the SF activity of companion galaxies. 
Star-forming companions often contribute to an enhanced supply of molecular gas (\citealt{2019MNRAS.485.1320M}; \citealt{2025ApJ...980..157H}) and increase the rate of interstellar medium (ISM) collisions (\citealt{2016ApJS..222...16C}; \citealt{2018ApJ...868..132P}; \citealt{2018ApJS..237....2Z}), both of which elevate the SFE and sSFRs in the interacting systems. In contrast, passive companions tend to retain stable hot gas halos (\citealt{2018MNRAS.473..538C}), which, when close, can suppress SF in the target galaxy by preventing the accretion of cold gas and even stripping existing gas reservoirs (\citealt{2013MNRAS.432..336W}; \citealt{2019MNRAS.486.5184R}; \citealt{2021PASA...38...35C}).
As a result, interaction-driven SF can be either enhanced or suppressed depending on the companion's SF activity. 
Furthermore, \href{https://iopscience.iop.org/article/10.3847/1538-4357/ad2063}{Paper V} revealed that the hydrodynamical response of galaxies during interactions depends on the relative orientation of their spin axes: SF enhancement is more pronounced when the galaxies are well aligned.

Building upon these findings, the present study (Paper VI) provides a coherent framework in which stellar bars play a critical role in modulating the hydrodynamical impact of galaxy interactions on central SF. 
Bars can increase or decrease the SF in the target galaxy, depending on how active SF is in the companion galaxy.
This dual role offers a natural explanation for the seemingly conflicting results reported in previous studies of barred galaxies (e.g., \citealt{2020MNRAS.499.1406L} ; \citealt{2020MNRAS.499.1116F} ; \citealt{2020A&A...644A..79G} ; \citealt{2020ApJ...893...19W}; \citealt{2023ApJ...949...91Z} ; \citealt{2025A&A...696A.118R}). 
The SF activity of companion galaxies can serve as a tracer of their cold gas content, especially molecular gas, which is the direct fuel for SF. 
Observational studies established a strong correlation between SF activities and molecular gas masses (e.g., \citealt{1998ApJ...498..541K}; \citealt{2011MNRAS.415...32S}). 
Specifically, \citealt{2015MNRAS.449.3719S} found a positive correlation between HI gas fraction and SFR enhancement at $\sim$\,2.5\,$\sigma$ level from 34 pair galaxies relative to the isolated sample.  
Star-forming companions imply the presence of substantial cold gas reservoirs not only within themselves but also suggest gas-rich interaction environments. 
In such cases, stellar bars tend to be associated with elevated central sSFRs and younger stellar populations (suggesting bars trigger starbursts; e.g., \citealt{1996A&A...313...13H}; \citealt{2012MNRAS.423.3486W}; \citealt{2019MNRAS.484.5192C}).
Conversely, a passive, quiescent companion implies an overall gas-poor environment, where bars are more likely to stabilize the disk and suppress further inflow, facilitating morphological quenching (e.g., \citealt{2012MNRAS.424.2180M}; \citealt{2021MNRAS.507.4389G}).
Our results unify the two distinct perspectives by showing that bars can induce either starbursts or quenching depending on gas environments.
Bars drive central SF when ample gas is available, while they inhibit it in environments lacking fresh gas supply. 
Without distinguishing between these two modes, past studies likely combined barred galaxies from both regimes (along with isolated galaxies), thereby masking the true dual effect of bars on SF.

\subsection{Bars in Gas-rich Interactions: Catalysts for Star Formation}

SF enhancement in galaxy interactions is commonly attributed to a combination of increased gas availability and elevated SFE.
Observationally, numerous barred galaxies in interacting systems exhibit extended HI bridges connecting to nearby companions, suggesting potential cold gas inflow. 
Examples include NGC 3395/3396 (\citealt{1999MNRAS.308..364C}; \citealt{2024MNRAS.532.1744Y}), NGC 6821/IC 4970 (\citealt{2007A&A...464..155H}), NGC 4490/4485 (\citealt{2010ApJ...723.1375R}), NGC 4105/4106 (\citealt{2021MNRAS.501.3750R}), and NGC 7582, which connects to NGC 7590 and NGC 7599 via a prominent HI bridge (\citealt{2005A&A...429L...5D}). 
An extreme case is NGC 6221, where an HI bridge spanning over $100\,\textrm{kpc}\,h^{-1}$ links it to NGC 6215 (\citealt{2004MNRAS.348.1255K}). 
These observations imply that gas-rich companions may directly supply fuel for bar-driven central starbursts.


The rotating potential of a stellar bar exerts gravitational torques on the ISM, efficiently removing angular momentum and driving gas inward toward the galactic center (\citealt{1965MNRAS.130..183F}; \citealt{2005ApJ...620..197A}; \citealt{2009A&A...496...85G}; \citealt{2020A&A...644A..38D}; \citealt{2022A&A...666A.175Y}). 
As gas flows along the bar, it encounters large-scale shocks---often traced by dust lanes---that dissipate kinetic energy and further funnel gas into the central kpc region. 
Recent observations suggested these bar-driven inflows as a dominant mechanism for fueling central SF, surpassing the role of external gas accretion. 
\citet{2010ApJ...723.1255R} found that central metallicity dilution is consistent with gas inflow.
\citet{2019ApJ...881..119P}, using MaNGA data, identified steepened sSFR gradients after first pericentric passage in gas-rich pairs. 
\citet{2025ApJ...982..130P} showed that such interactions can reform disks with flattened metallicity profiles. 
Most recently, \citet{2025MNRAS.538..327W}, using HI data from the FAST Extended Atlas of Selected Targets Survey (FEASTS), found that tidal stripping mainly affects the lower-mass companion, while the target galaxy experiences both accretion and depletion of gas.

Simulations also showed that bar-induced gravitational torques and shocks efficiently funnel gas inward, building central concentrations and triggering nuclear starbursts or AGN activity (\citealt{1983MNRAS.205.1009N}; \citealt{1991ApJ...370L..65B}; \citealt{1996ApJ...464..641M}; \citealt{2008MNRAS.384..386C}). 
In gas-rich major mergers, this can yield ultra-luminous infrared galaxies (ULIRGs).
\citet{2025ApJ...981..156M} confirmed that collisions between molecular clouds occur more frequently due to kpc-scale inflows at the bar ends in NGC 3627, while the overall suppression of SF depends on variations in the SFE. 
These findings demonstrate that, while direct gas accretion from companions plays a limited role, gas-rich interactions and bar dynamics jointly redistribute existing gas toward the center---establishing conditions favorable for bar-driven SF enhancement.
In gas-rich systems, this synergy positions stellar bars as \textit{catalysts} of interaction-driven central starbursts.

\subsection{Bars in Gas-poor Interactions: Drivers of Quenching}

Bar incidence is higher in massive, passive, and gas-poor disks (\citealt{2016A&A...595A..63V}; \citealt{2020MNRAS.499.1116F}), consistent with two complementary pictures: ($a$) high gas fractions can delay or weaken bar formation (\citealt{2012MNRAS.424.2180M}; \citealt{2013MNRAS.429.1949A}); 
and ($b$) once established in gas-poor systems, bars contribute to quenching by accelerating fuel consumption and stabilizing the disk (\citealt{2018A&A...609A..60K}; \citealt{2019A&A...628A..24G}; \citealt{2020MNRAS.499.1116F}; \citealt{2021A&A...651A.107G}).
Using Galaxy Zoo: DECaLS morphologies, \citet{2021MNRAS.507.4389G} found that both weak and strong bars are more common in quiescent galaxies, with strong bars linked to elevated central SFRs and shorter gas-depletion timescales—consistent with rapid consumption preceding quenching.
Observational analyses from PHANGS-MUSE and PHANGS-ALMA confirmed that bars induce strong velocity shear and shocks along the bar, which can suppress SF by stabilizing the disk against collapse (\citealt{2024ApJ...968...87K}).

Although galaxy–galaxy interactions often enhance SF in gas-rich encounters—where bars can efficiently funnel fresh gas and elevate central SFRs (e.g., \citealt{2007A&A...468...61D}; \citealt{2019ApJ...881..119P}; \href{https://iopscience.iop.org/article/10.3847/1538-4357/ab3401}{Paper I}; \href{https://iopscience.iop.org/article/10.3847/1538-4357/ad2063}{Paper V})—the outcome can depend on sustained gas supply and on timescale 
(\citealt{2008A&A...492...31D}; \citealt{2010MNRAS.404..590L}; \citealt{2010MNRAS.407.2091G}; \citealt{2024ApJ...977..116H}). 
In systems lacking continuous replenishment, especially with passive, gas-deficient companions, non-axisymmetric bar torques drive a short-lived nuclear boost (\citealt{2013MNRAS.430.1901H}; \citealt{2017MNRAS.465.1934F}) and then exhaust the available fuel on $\sim10^{8-9}$ yr. 
The resulting central mass concentration, together with strong shocks and velocity shear along the bar, stabilizes the disk (higher effective Toomre $Q$) and suppresses new SF despite the presence of a prominent bar (\citealt{1998A&A...337..671R}; \citealt{2018A&A...609A..60K}; \citealt{2024ApJ...968...87K}). 
Observations support this replenishment-limited pathway.
\citet{2025A&A...696A.118R} found that barred galaxies often exhibit extended $NUV-r$ color gradients and redder disks, indicative of older stellar populations and halted SF beyond the nucleus, likely due to exhausted central gas and lack of inflow. 
HI-based studies supported this picture. 
\citet{2020MNRAS.492.4697N} reported that barred galaxies with the lowest HI content show prominent central HI holes and minimal SF, while gas-rich barred systems remain actively star-forming. 
Similarly, in a multi-wavelength analysis, \citet{2020A&A...644A..79G} found that only NGC 2903, which retains an outer HI reservoir, sustains SF; the other three galaxies (NGC 3351, NGC 4579, and NGC 4725), all gas-poor, exhibit no ongoing SF. 
These findings suggest that in the absence of external gas supply---particularly in interactions with passive, gas-deficient companions---stellar bars act as effective \textit{inhibitors} of SF.

The observed trend, in which strongly barred galaxies are more likely to have passive companions (Figure~\ref{fig:3}), also suggests a link between bar evolution and environmental SF quenching.
Passive galaxies are more common in denser regions, where environmental processes such as ram-pressure stripping or starvation suppress SF (e.g., \citealt{2016A&A...596A..11B}; \citealt{2019MNRAS.483.2851S}; \citealt{2022A&A...666A.141M}; \citealt{2024MNRAS.534.3974K}). 
The same environments exhibit higher bar fractions, especially in massive systems (e.g., \citealt{2011MNRAS.411.2026M}; \citealt{2020ApJ...893..117Y}). 
Consequently, the association between strong bars and quenched neighbors may arise from a combination of internal dynamical stability, limited external gas supply, and environmental quenching. 
These results imply that strong bars preferentially emerge or persist in gas-poor interactions, reinforcing their role in secular evolution and the suppression of SF (e.g., \citealt{2013ApJ...779..162C}; \citealt{2018MNRAS.473.4731K}).

\section{Conclusions}\label{sec:sum}

We investigate how the bar of a target galaxy and the star-forming activity of its companion galaxy jointly regulate SF of the target galaxy in an interacting pair, using a large sample of barred systems ($0.02 < z < 0.08$) drawn from the DESI-LS.
Our study reveals that stellar bars have a dual role in the regulation of SF in interacting systems. 
Bars can either ignite or extinguish central SF---specifically, depending on whether or not a close companion can provide fresh gas. 
Our main findings are summarized as follows.

\begin{enumerate}
    \item 
    \textbf{Enhanced Bimodality of the SF Distribution:} Barred galaxies in pairs show a bimodal sSFR distribution. 
    Relative to unbarred pairs (which largely mirror isolated galaxies), barred galaxies in pairs populate both the high-SF tail and the low-SF tail of the distribution. 
    Approximately one-third of strongly barred galaxies in pairs are quenched ($\textrm{Log}(\rm sSFR) < -11 \hspace{1mm} yr^{-1}$), while two-third experience SF boosts well above the level for the control sample.
    \item 
    \textbf{Dual Role of Bars:} Bars exhibit a two-faced influence on interaction-driven SF. 
    In strongly barred galaxies with gas-rich, star-forming companions, interactions lead to pronounced central starbursts (sSFR enhancements up to 2$-$3$\times$ the norm). 
    Conversely, in strongly barred galaxies with gas-poor, passive companions, interactions result in SF quenching. 
    \item 
    \textbf{Interaction Proximity and Companion Properties:} The impact of an interaction on SF is stronger at smaller separations, but only if the companion is star-forming. 
    Close encounters with star-forming companions drive significant SF enhancement, especially in barred galaxies. 
    Close encounters with passive companions yield little or no SF enhancement, and in barred cases can even suppress SF. 
    This recalls the importance of the companion gas content in determining the outcome during interactions. 
    \item 
    \textbf{Bars and Interaction Influence:} We confirm that bars are not significantly more common in currently interacting galaxies compared to more isolated galaxies; i.e., the bar fraction does not rise toward smaller separations. 
    This suggests that bars in our sample are likely in place before the current interaction, or that interactions can both create and destroy bars with roughly equal likelihood. 
    Regardless, the presence of a bar is a critical factor that governs how an interaction affects the galaxy's SF. 
\end{enumerate}

Our results of bar's dual role in regulating SF in interacting systems propose that bars serve as a key internal mechanism linking kpc-scale disk dynamics to external environmental influence.
In gas-rich interactions---typically involving star-forming companions---bars efficiently channel gas into the central regions, fueling enhanced SF. 
In contrast, when the companion is passive and the interaction lacks a fresh gas supply, bars appear to accelerate gas depletion and contribute to central quenching.
This highlights the critical importance of simultaneously considering both internal structures and external companion properties to fully understand the mechanisms regulating SF during galaxy interactions.

Future progress will require spatially resolved spectroscopy from integral field unit surveys, such as MaNGA and JWST, to trace internal gas transfer and determine whether gas from companion galaxies directly fuels or suppresses SF in target galaxies.
In parallel, expanded statistical samples with deeper imaging and spectroscopic coverage will be essential to assess how companion properties---such as gas content, morphology, and kinematics---modulate bar-driven evolution across a range of environments and redshifts. 
On the simulations, cosmological models can incorporate both interaction-driven dynamics and secular bar evolution to reproduce the dual modes of SF regulation observed in this work.


\begin{acknowledgments}
This work was supported by the faculty research fund of Sejong University in 2025.
S.-J.Y. acknowledges support from the Mid-career Researcher Programs (RS-2024-00344283) and the Basic Science Research Program (RS-2022-NR070872) through the National Research Foundation (NRF) of Korea.
The data in this paper are the result of the efforts of the Galaxy Zoo volunteers, without whom none of this work would be possible. Their efforts are individually acknowledged at \url{http://authors.galaxyzoo.org}.
Funding for the Sloan Digital Sky Survey V has been provided by the Alfred P. Sloan Foundation, the Heising-Simons Foundation, the National Science Foundation, and the Participating Institutions. SDSS acknowledges support and resources from the Center for High-Performance Computing at the University of Utah. SDSS telescopes are located at Apache Point Observatory, funded by the Astrophysical Research Consortium and operated by New Mexico State University, and at Las Campanas Observatory, operated by the Carnegie Institution for Science. The SDSS web site is \url{http://www.sdss.org}.
SDSS is managed by the Astrophysical Research Consortium for the Participating Institutions of the SDSS Collaboration, including Caltech, The Carnegie Institution for Science, Chilean National Time Allocation Committee (CNTAC) ratified researchers, The Flatiron Institute, the Gotham Participation Group, Harvard University, Heidelberg University, The Johns Hopkins University, L'Ecole polytechnique f\'{e}d\'{e}rale de Lausanne (EPFL), Leibniz-Institut f\"{u}r Astrophysik Potsdam (AIP), Max-Planck-Institut f\"{u}r Astronomie (MPIA Heidelberg), Max-Planck-Institut f\"{u}r Extraterrestrische Physik (MPE), Nanjing University, National Astronomical Observatories of China (NAOC), New Mexico State University, The Ohio State University, Pennsylvania State University, Smithsonian Astrophysical Observatory, Space Telescope Science Institute (STScI), the Stellar Astrophysics Participation Group, Universidad Nacional Aut\'{o}noma de M\'{e}xico, University of Arizona, University of Colorado Boulder, University of Illinois at Urbana-Champaign, University of Toronto, University of Utah, University of Virginia, Yale University, and Yunnan University.
The Legacy Surveys consist of three individual and complementary projects: the Dark Energy Camera Legacy Survey (DECaLS; Proposal ID \#2014B-0404; PIs: David Schlegel and Arjun Dey), the Beijing-Arizona Sky Survey (BASS; NOAO Prop. ID \#2015A-0801; PIs: Zhou Xu and Xiaohui Fan), and the Mayall z-band Legacy Survey (MzLS; Prop. ID \#2016A-0453; PI: Arjun Dey). DECaLS, BASS and MzLS together include data obtained, respectively, at the Blanco telescope, Cerro Tololo Inter-American Observatory, NSF’s NOIRLab; the Bok telescope, Steward Observatory, University of Arizona; and the Mayall telescope, Kitt Peak National Observatory, NOIRLab. Pipeline processing and analyses of the data were supported by NOIRLab and the Lawrence Berkeley National Laboratory (LBNL). The Legacy Surveys project is honored to be permitted to conduct astronomical research on Iolkam Du’ag (Kitt Peak), a mountain with particular significance to the Tohono O’odham Nation.
NOIRLab is operated by the Association of Universities for Research in Astronomy (AURA) under a cooperative agreement with the National Science Foundation. LBNL is managed by the Regents of the University of California under contract to the U.S. Department of Energy.
This project used data obtained with the Dark Energy Camera (DECam), which was constructed by the Dark Energy Survey (DES) collaboration. Funding for the DES Projects has been provided by the U.S. Department of Energy, the U.S. National Science Foundation, the Ministry of Science and Education of Spain, the Science and Technology Facilities Council of the United Kingdom, the Higher Education Funding Council for England, the National Center for Supercomputing Applications at the University of Illinois at Urbana-Champaign, the Kavli Institute of Cosmological Physics at the University of Chicago, the Center for Cosmology and Astro-Particle Physics at the Ohio State University, the Mitchell Institute for Fundamental Physics and Astronomy at Texas A\&M University, Financiadora de Estudos e Projetos, Fundacao Carlos Chagas Filho de Amparo, Financiadora de Estudos e Projetos, Fundacao Carlos Chagas Filho de Amparo a Pesquisa do Estado do Rio de Janeiro, Conselho Nacional de Desenvolvimento Cientifico e Tecnologico and the Ministerio da Ciencia, Tecnologia e Inovacao, the Deutsche Forschungsgemeinschaft and the Collaborating Institutions in the Dark Energy Survey. 
The Collaborating Institutions are Argonne National Laboratory, the University of California at Santa Cruz, the University of Cambridge, Centro de Investigaciones Energeticas, Medioambientales y Tecnologicas-Madrid, the University of Chicago, University College London, the DES-Brazil Consortium, the University of Edinburgh, the Eidgenossische Technische Hochschule (ETH) Zurich, Fermi National Accelerator Laboratory, the University of Illinois at Urbana-Champaign, the Institut de Ciencies de l’Espai (IEEC/CSIC), the Institut de Fisica d’Altes Energies, Lawrence Berkeley National Laboratory, the Ludwig Maximilians Universitat Munchen and the associated Excellence Cluster Universe, the University of Michigan, NSF’s NOIRLab, the University of Nottingham, the Ohio State University, the University of Pennsylvania, the University of Portsmouth, SLAC National Accelerator Laboratory, Stanford University, the University of Sussex, and Texas A\&M University.
BASS is a key project of the Telescope Access Program (TAP), which has been funded by the National Astronomical Observatories of China, the Chinese Academy of Sciences (the Strategic Priority Research Program “The Emergence of Cosmological Structures” Grant \# XDB09000000), and the Special Fund for Astronomy from the Ministry of Finance. The BASS is also supported by the External Cooperation Program of Chinese Academy of Sciences (Grant \# 114A11KYSB20160057), and Chinese National Natural Science Foundation (Grant \# 12120101003, \# 11433005).
The Legacy Survey team makes use of data products from the Near-Earth Object Wide-field Infrared Survey Explorer (NEOWISE), which is a project of the Jet Propulsion Laboratory/California Institute of Technology. NEOWISE is funded by the National Aeronautics and Space Administration.
The Legacy Surveys imaging of the DESI footprint is supported by the Director, Office of Science, Office of High Energy Physics of the U.S. Department of Energy under Contract No. DE-AC02-05CH1123, by the National Energy Research Scientific Computing Center, a DOE Office of Science User Facility under the same contract; and by the U.S. National Science Foundation, Division of Astronomical Sciences under Contract No. AST-0950945 to NOAO.
The Photometric Redshifts for the Legacy Surveys (PRLS) catalog used in this paper was produced thanks to funding from the U.S. Department of Energy Office of Science, Office of High Energy Physics via grant DE-SC0007914.

\end{acknowledgments}

\end{document}